\documentclass[runningheads,a4paper,journal]{ieeetran} 

\usepackage{amssymb}
\usepackage{graphicx}

\usepackage{url}
\urldef{\mailsa}\path|{henk.d.l.hollmann, paul.gorissen,|
\urldef{\mailsc}\path|ronald.rietman, sebastiaan.de.hoogh, ludo.tolhuizen}@philips.com|

\usepackage{amsmath}
\usepackage{algorithm}
\usepackage{algorithmicx} 
\usepackage{algpseudocode}

\usepackage{amssymb}
\usepackage{amsmath} 
\usepackage{ntheorem}
\def\mymedskip{\vskip\medskipamount}

\def\mymedbreak{\par \ifdim\lastskip<\medskipamount
  \removelastskip \penalty-100 \mymedskip \fi}
\def\myaftermedspace{\par \ifdim\lastskip<\medskipamount
  \removelastskip \penalty55\mymedskip\fi}
\newcommand{\eop}{{\unskip\nobreak\hfil\penalty50
          \hskip2em\hbox{}\nobreak\hfil$\Box$
          \parfillskip=0pt \finalhyphendemerits=0 \par}}
\newenvironment{proof}%
{\mymedbreak{\noindent\bf Proof.\enspace}}{\eop\myaftermedspace}
\newenvironment{proofn}[1]%
{\mymedbreak{\noindent\bf Proof #1.\enspace}}{\eop\myaftermedspace}
{\mymedbreak{\noindent\bf Proof of Theorem #1:\enspace}}{\eop\myaftermedspace}
{\mymedbreak\noindent{\bf Remark:}%
\enspace\rm}%
{\myaftermedspace}
\newtheorem{teor}{Theorem}[section]
\newtheorem{defi}[teor]{Definition}
\newtheorem{fact}[teor]{Fact}
\newtheorem{problem}{Problem}
\newtheorem*{problemnn}{Problem}
\newtheorem{exercise}{Exercise}
\newtheorem{examp}[teor]{Example}
\newtheorem{lem}[teor]{Lemma}
\newtheorem{cor}[teor]{Corollary}
\newtheorem{con}[teor]{Conjecture}
\newtheorem{prop}[teor]{Proposition}
\newtheorem{rem}[teor]{Remark}
\newtheorem{alg}[teor]{Algorithm}
\newcommand{\beq}{\begin{equation}}
\newcommand{\eeq}{\end{equation}}
\newcommand{\beql}[1]{\begin{equation} \label{#1}}
\newcommand{\eeql}{\end{equation}}
\newcommand{\beqa}{\begin{eqnarray*}}
\newcommand{\eeqa}{\end{eqnarray*}}
\newcommand{\beqal}[1]{\begin{eqnarray} \label{#1}}
\newcommand{\eeqal}{\end{eqnarray}}
\newcommand{\beqan}{\begin{eqnarray}}
\newcommand{\eeqan}{\end{eqnarray}}
\newcommand{\bpf}{\begin{proof}}
\newcommand{\epf}{\end{proof}}
\newcommand{\bpfn}[1]{\begin{proofn}{#1}}
\newcommand{\epfn}{\end{proofn}}
\newcommand{\ben}{\begin{enumerate}}
\newcommand{\een}{\end{enumerate}}
\newcommand{\bit}{\begin{itemize}}
\newcommand{\eit}{\end{itemize}}

\newcommand{\bab}{\begin{abstract}}
\newcommand{\eab}{\end{abstract}}
\newcommand{\bke}{\begin{keywords}}
\newcommand{\eke}{\end{keywords}}

\newcommand{\btm}[1]{\begin{teor} \label{#1}}
\newcommand{\etm}{\end{teor}}
\newcommand{\btmn}[2]{\begin{teor}[#1] \label{#2}}
\newcommand{\etmn}{\end{teor}}
\newcommand{\ble}[1]{\begin{lem} \label{#1}}
\newcommand{\ele}{\end{lem}}
\newcommand{\bLe}[1]{\begin{Lemma} \label{#1}}
\newcommand{\eLe}{\end{Lemma}}
\newcommand{\bpn}[1]{\begin{prop} \label{#1}}
\newcommand{\epn}{\end{prop}}
\newcommand{\bex}[1]{\begin{examp} \label{#1}}
\newcommand{\eex}{\hfill$\Box$\end{examp}}
\newcommand{\bde}[1]{\begin{defi} \label{#1}}
\newcommand{\ede}{\end{defi}}
\newcommand{\bco}[1]{\begin{cor} \label{#1}}
\newcommand{\eco}{\end{cor}}
\newcommand{\bcorn}[2]{\begin{cor}[#1] \label{#1}}
\newcommand{\ecorn}{\end{cor}}
\newcommand{\bcon}[1]{\begin{con} \label{#1}}
\newcommand{\econ}{\end{con}}
\newcommand{\bfact}[1]{\begin{fact} \label{#1}}
\newcommand{\efact}{\end{fact}}
\newcommand{\bpr}[1]{\begin{problem} \label{#1}}
\newcommand{\epr}{\end{problem}}
\newcommand{\bprnn}[1]{\begin{problemnn} \label{#1}}
\newcommand{\eprnn}{\end{problemnn}}
\newcommand{\bprn}[2]{\begin{problem}[#1] \label{#2}}
\newcommand{\eprn}{\end{problem}}
\newcommand{\bexer}[1]{\begin{exercise} \label{#1}}
\newcommand{\eexer}{\end{exercise}}
\newcommand{\bre}[1]{\begin{rem} \label{#1}}
\newcommand{\ere}{\end{rem}}
\newcommand{\balg}[1]{\begin{alg} \label{#1}}
\newcommand{\ealg}{\end{alg}}
\newcount\tpcnt
\newenvironment{tproblem}{%
  \global\advance\tpcnt1%
  \goodbreak\medskip\par\noindent\textbf{Problem~\the\tpcnt.}~}%
{%
  \goodbreak
}

\newenvironment{Solution}[1][]{%
  \goodbreak\smallskip\par\noindent\textbf{Solution{\if#1\empty\else~#1\fi}.}~}%
{%
  \goodbreak
}

\newcommand{\cA}{{\cal A}}
\newcommand{\cB}{{\cal B}}

\newcommand{\cI}{{\cal I}}

\newcommand{\cM}{{\cal M}}

\newcommand{\cR}{{\cal R}}
\newcommand{\cS}{{\cal S}}

\newcommand{\gw}{\omega}

\newcommand{\gvf}{\varphi}

\newcommand{\ga}{\alpha}
\newcommand{\gb}{\beta}

\newcommand{\gd}{\delta}

\newcommand{\gre}{\epsilon}

\newcommand{\Tm}[1]{Theorem~\protect\ref{#1}}

\newcommand{\Pn}[1]{Proposition~\protect\ref{#1}}

\newcommand{\Co}[1]{Corollary~\protect\ref{#1}}

\newcommand{\Sec}[1]{Section~\protect\ref{#1}}

\newcounter{question_number}
\newenvironment{question}{\addtocounter{question_number}{1}\noindent{\bf Question \arabic{question_number}}}{\myaftermedspace}

\newenvironment{solution}{\noindent {\bf Solution:} \enspace}{\eop\myaftermedspace}

\newenvironment{hint}{\noindent {\bf Hint:} \enspace}{\eop\myaftermedspace}

\newenvironment{multisolution}[1]{\noindent {\bf Solution #1:} \enspace}{\eop\myaftermedspace}

\newcommand{\bqu}{\begin{question}}
\newcommand{\equ}{\end{question}}
\newcommand{\bs}{\begin{solution}}
\newcommand{\es}{\end{solution}}
\newcommand{\bh}{\begin{hint}}
\newcommand{\eh}{\end{hint}}
\newcommand{\bms}[1]{\begin{multisolution}{#1}}
\newcommand{\ems}{\end{multisolution}}
\newcommand{\bquo}{\begin{quote}}
\newcommand{\equo}{\end{quote}}

\newcommand{\btp}{\begin{tproblem}}
\newcommand{\etp}{\end{tproblem}}
\newcommand{\bts}{\begin{Solution}}
\newcommand{\ets}{\end{Solution}}

\newcommand{\lcm}{{\rm lcm}}

\newcounter{penumi}
\newenvironment{pit}{%
\begin{list}{(\roman{penumi})}{\usecounter{penumi}\setlength{\labelwidth}{1cm}\setlength{\itemindent}{0pt}\setlength{\topsep}{0pt}\setlength{\parsep}{0pt}\setlength{\partopsep}{0pt}\setlength{\itemsep}{0pt}}
} 
{\end{list}}
\newcommand{\bpit}{\begin{pit}}
\newcommand{\epit}{\end{pit}}

\newcommand{\xhat}{{\hat{x}}}

\newcommand{\red}[2]{|#1|_{#2}}

\newcommand{\amp}{{\&}}
\newcommand{\IPS}{IP\amp S}
\newcommand{\To}{\textbf{to}\ }
\newcommand{\resi}{residue\ }

\begin{document} 

\title{%
A Multi-layer Recursive Residue Number System} 
\date{\today} 
\author{%
Henk D.L.~Hollmann,
Ronald Rietman,
Sebastiaan de Hoogh,
Ludo~M.G.M.~Tolhuizen,~\IEEEmembership{Senior Member,~IEEE},
and Paul Gorissen%
\thanks{Henk D.L.~Hollmann and Paul~Gorissen are with Philips \IPS, Eindhoven, the Netherlands;
Ronald~Rietman,
Sebastiaan~de~Hoogh, and
Ludo~Tolhuizen are with Philips Research, Eindhoven, the Netherlands}%
\thanks{Email: \{henk.d.l.hollmann, ronald.rietman, sebastiaan.de.hoogh, ludo.tolhuizen, paul.gorissen\}@philips.com}%

%
%
%


}

%
%


\maketitle 
\pagestyle{plain}
\begin{abstract}
We present a method to increase the dynamical range of a Residue Number System (RNS) by adding virtual RNS layers on top of the original RNS, where the  required modular arithmetic for a modulus on any non-bottom layer is implemented by means of an RNS Montgomery multiplication algorithm that uses  the RNS on the layer below. As a result, the actual arithmetic is deferred to the bottom layer. The multiplication algorithm that we use is based on an algorithm by Bajard and Imbert, extended to work with pseudo-residues (remainders with a larger range than the modulus). The resulting Recursive Residue Number System (RRNS) can be used to implement modular addition, multiplication, and multiply-and-accumulate for very large (2000+ bits) moduli, using only modular operations for small (for example 8-bits) moduli. A hardware implementation of this method allows for massive parallelization. 

Our method can be applied in cryptographic algorithms such as RSA to realize modular exponentiation with a large (2048-bit, or even 4096-bit) modulus. Due to the use of full RNS Montgomery algorithms, the system does not involve any carries,  therefore cryptographic attacks that exploit carries cannot be applied.
\end{abstract} 


%
%

%
\section{Introduction}
A Residue Number System (RNS) allows parallel (hence carry-free) addition,
subtraction and multiplication on the representing residues of the operands, thus
promising large gains in speed compared to arithmetic on numbers in conventional representation. The high speed and low power-consumption of RNS due
to the absense of carries makes it attractive for use in embedded processors,
such as those found in mobile devices \cite{op-rns}. RNS is also useful for applications
to fault-tolerant computing \cite{op-rns}. 
%
Unfortunately, other operations such as division, magnitude comparison, and residue-to-digital conversion are more difficult for numbers in RNS representation, therefore the use of RNS is mainly interesting in applications where most of the required operations are addition/subtraction and multiplication, with only rarely a need for conversion to other (digital) representations. 
Typical applications for RNS can be found in Digital Filtering, image
and speech processing, and cryptography.
The classical reference for RNS is \cite{st-ra}; for additional background, see for example  \cite{op-rns,m-rns,p-ca}.

Much research has been done in the implementation of modular multiplication methods such as Montgomery multiplication \cite{montgomery1985modular} and Barrett multiplication \cite{barrett1986} in RNS arithmetic. An interesting method for Montgomery multiplication in an RNS is the {\em Bajard-Imbert algorithm\/} described in~\cite{bi-rns}.  This is a full RNS method, thus inherently carry-free.

To meet the present-day or near-future security requirements, moduli of 2048 or even 4096 bits in RSA are needed.
%
To realize RNS systems with a dynamical range (the range of values that the RNS can represent) that is sufficiently large to implement the Bajard-Imbert RNS algorithm for 2048-bits moduli, we would need rather large moduli, since there simply are not enough small moduli available. For example, employing only 8-bit moduli, the largest attainable dynamical range is $\lcm(2,3,\ldots, 256)$, a 363-bits number.  The conventional solution is to employ moduli of a special type, for example  of the form $2^n-c$ for small~$c$, see, e.g.,   \cite{bdem-rns,cnpq-rsa}. However, the direct implementation of the modular arithmetic for such moduli would not be carry-free. It has been shown in \cite{Fouque2008}
and \cite{Gebotys-HMACCarryAttack} how to exploit the leakage of carries through side-channel analysis in the context of RSA and HMAC, respectively.
The basic idea is that if random integers $R_i$ are added to a fixed number $S$, the probability distribution of carry values over different values of $R_i$ 
depends on $S$. Indeed, if $S$ is  large (or small), carry values will often be equal; if $S$ has an intermediate value,
the carry value distribution is much less skew.  A similar attack can be mounted in the context of whitebox cryptography  \cite{Chow-wb}:
if a table has two encoded input values
and produces an encoded carry value, then  by fixing one input and observing the distribution of the encoded carry value if 
the other input runs through all values, an attacker
can obtain information on the fixed input. In this way, the attacker can obtain much information on the input encodings.

This paper presents a methods to implement multi-layer RNS systems with virtual unlimited dynamical range while still employing only modular arithmetic for small moduli. 
The idea behind the method is to implement the modular arithmetic for the RNS moduli in each (non-bottom) layer using an RNS-based, Bajard-Imbert-type algorithm that employs the RNS on the layer below. As a consequence, all the modular arithmetic is deferred to the bottom RNS layer, which consists of small moduli only; moreover, all the arithmetic methods are truly carry-free. The Bajard-Imbert algorithm implements a Montgomery multiplication modulo a modulus~$N$, and delivers the result in the form of a {\em pseudo-residue\/} modulo~$N$,  a number with a range larger than the minimum required $N$ values. 
The algorithm that we present can be seen as a slightly improved Bajard-Imbert algorithm that can also operate with such pseudo-residues. 
In contrast to known multi-layer (also called hierarchical) RNS systems that only do modular arithmetic on the bottom layer \cite{Y-hier}, or that use moduli on a higher layer that are the products of some of the moduli on the lower layer  \cite{sa-rns,t-hier}, our method applies non-trivial modular arithmetic in all layers.  

Our multi-layer RNS method allows for massive parallelization, with one, or even a multitude, of processors per bottom modulus, so the method may also be suitable for high-speed applications. 

The contents of this paper are as follows. In the next section, we present some background concerning Residue Number Systems and Montgomery multiplication techniques, and we outline the Bajard-Imbert  algorithm. We introduce pseudo-residues in the form as used in this paper, and we indicate  how to adapt the base-extension method used in the Bajard-Imbert algorithm so that it can operate with pseudo-residues. 
In \Sec{LS-mmrns}, we first discuss the requirements on the RNS moduli in the current top-lay  that are needed when adding the next RNS layer. Then we describe in detail our new Bajard-Imbert-type full RNS algorithm, including a motivation for each step. We then discuss the additional requirements on the redundant modulus. 
In \Sec{LS-B} we describe the precise conditions for correctness of our algorithm, in the form of various bounds that have to be satisfied. We also indicate several ways to improve our algorithm. Then in \Sec{LS-comp}, we present complexity estimates, which we use to indicate how to design an efficient system.
Finally, in \Sec{LS-ex}, we present an example of a two-layer RNS that implement modular arithmetic for 2048-bits moduli by adding the desired RNS modulus on top,  in a third layer. This system employs a bottom layer RNS consiting of  19 small,  8-bit moduli  (the second, ``virtual'' layer consists of 64 moduli of 66 bits each). As a result, all the arithmetic can, for example,  be done by 8-bit by 8-bit table look-up; here one modular multiplication would take about 160,000 table look-ups.  
We note that in order to change  the RSA modulus $N$, only a limited number of constants in the algorithm have to be adapted. This update need not be computed in a secure manner and hence can be done quickly.

Our (non-parallelized) C++  program  implementing this two-layer RNS  exponentiation algorithm with table lookup arithmetic takes approximately .3 second to do a 500-bit modular exponentiation for a 2048-bit modulus on an HP Elitebook Folio 9470m  laptop.
We conclude that the security wish to remove all carries in the arithmetic can be satisfied with 
an implementation that still operates at an acceptable speed.
%
%
In another example, we show that a multi-layer RNS with a very large dynamical range can already be created starting with a  bottom layer consisting of five 4-bit moduli.
We end the paper by presenting some conclusions.


%
\section{Background and notation}
\subsection{\label{LSS-CRT}Residue Number Systems}
A {\em Residue Number System\/} or RNS  represents an integer~$x$ with respect to a {\em base\/} $\cB=(M_1, \ldots, M_k)$, consisting of positive mutually co-prime integers, by a sequence of  integers $(x_1, \ldots, x_k)$, where 
$ x\equiv x_i \bmod M_i$ for all~$i$.
More general, given integer constants $H_1, \ldots, H_k$ with $\gcd(H_i,M_i)=1$ for all~$i$, we will refer to a $k$-tuple $\ga=(\ga_1, \ldots, \ga_k)$ for which $\ga_i\equiv xH_i^{-1}\bmod M_i$ 
for all $i$ as a {\em $H$-representation\/} of $x$ in the RNS $\cB$. 
The {\em dynamical range\/} of the RNS is defined to be the integer $M=M_1, \cdots M_k$. 
The constructive version of the Chinese Remainder Theorem (CRT) gives a method to recover the integer $x$ modulo $M$ from a $H$-representation $(\ga_1, \ldots, \ga_k)$ 
as
\[x\equiv \xi_1 (M/M_1)+\cdots+\xi_k(M/M_k) \bmod M,
\]
where $\xi_i \equiv \ga_iH_i(M/M_i)^{-1} \bmod M_i$. 
%
%


%
\subsection{\label{LSS-M}The Montgomery technique}
{\em Montgomery reduction\/} is a technique to replace the (difficult) division by the modulus~$N$ as required in modular reduction by an easier division by a suitably chosen integer $M$, the {\em Montgomery constant\/}, where $\gcd(N,M)=1$. (Typically, $M$ is chosen to be a power of 2.) Then  to reduce a given integer $h$ modulo $N$, with $0\leq h<MN$, we first compute the integer $u$ with  $0\leq u<M$ for which  $h+uN$ is divisible by $M$; now the Montgomery reduction of~$h$ is defined as  $z=(h+uN)/M$. Note that $0\leq z<2N$ and $z\equiv hM^{-1}\bmod N$.
The {\em  Montgomery multiplication\/} of two integers $x,y$ computes an integer $z$ for which $z\equiv xyM^{-1}\bmod N$ by letting $h=xy$ and then taking $z$ to be the Montgomery reduction of~$h$. We can
use Montgomery multiplication to compute modular multiplications as follows. A {\em Montgomery representation\/} of an integer $X$ is an integer $x$ for which $x\equiv XM\bmod N$. Then given Montgomery representations $x\equiv XM\bmod N$ and $y\equiv YM\bmod N$ of two integers $X,Y$, we can obtain a Montgomery representation $z\equiv ZM\bmod N$ of the product $Z\equiv XY \bmod N$ as  the Montgomery product of $x$ and $y$.  This works since 
%
\[z\equiv xyM^{-1} \equiv XYM   \equiv ZM\bmod N.\]
Note that if $M>4N$, then given two integers $x,y<2N$, we have $h=xy<4N^2<MN$, hence the Montgomery product $z$  of $x$ and $y$ again satisfies $z<2N$. As a consequence, the Montgomery technique is especially suitable for modular exponentiation. Indeed, note that an exponentiation $Y=X^e\bmod N$ can be computed by computing a Montgomery representation $x<2N$ of $X$, for example by a Montgomery multiplication of $X$ and $M^2 \bmod N$, followed by a sequence of Montgomery multiplications to compute a Montgomery representation $y$ of $Y$, where $y<2N$. Finally, $Y$ can be obtained by a Montgomery multiplication of $y$ by 1.  For further details, see for example  \cite{dhem,kak-acm,menezes1996handbook,montgomery1985modular}.

%
%
\subsection{Pseudo-residues and expansion bounds}
The {\em residue\/} of an integer $x$ with respect to a modulus $n$ is the unique integer $r$ for which $r\equiv x\bmod n$ and $0\leq r<n$; a {\em pseudo-residue\/} of~$x$ is a number of the form $r+qn$ with $q$ ``small'', in some sense. To make this precise, we introduce the following definition. 
\bde{LDpr}\rm Let $\cI$ be an interval of length 1 of the form $[-1+e, e)$ for some rational number~$e$,
and let $\gvf$ be a positive number. We  say that an integer~$r$ is a {\em pseudo-residue\/} modulo~$n$ for an integer $x$, with {\em expansion bound\/} $\gvf$ and {\em \resi interval\/}~$\cI$, if $r\equiv x\bmod n$ and $r\in \gvf n\cI$.
We use the shorthand notation $\cI^2=\cI\times \cI=\{ab\mid a,b\in \cI\}$. We will write $\red{x}{n}$ to denote the unique integer $r\equiv x \bmod n$ for which $r\in n\cI$. 
\ede

In this paper, we only consider \resi intervals of the form $\cI=[0,1)$, referred to as  {\em standard residues\/}, or of the form $\cI=[-1/2, 1/2)$, referred to as  {\em symmetric residues\/}. Note that ordinary residues correspond to the case of standard residues with expansion bound 1. 
The idea to integers other than $0, 1, \ldots, n-1$ to represent residues modulo $n$ has been used before, see for example \cite{p-ca}, \cite{ibra}, \cite{par-note}, \cite{par-sym}.
%

Typically, modular multiplication algorithms such as, for example,  Montgomery~\cite{montgomery1985modular}, Barrett~\cite{barrett1986}, or Quisquater~\cite{quisquater1992encoding} 
deliver the result in the form of a pseudo-residue. As seen in \Sec{LSS-M}, the output of a Montgomery multiplication is a pseudo-residue with (standard) expansion bound~2 when the inputs themselves are pseudo-residues with expansion bound~2.
As mentioned in the introduction,  for a multi-layer RNS method based on such a modular multiplication algorithm, we need an RNS implementation of the algorithm capable of handling inputs and outputs represented by pseudo-residues with given, fixed expansion. Our example method is based on the full RNS implementation of a Montgomery multiplication from~\cite{bi-rns}, with some modifications. That algorithm uses the base extension method from \cite{sk-fbe} that employs a redundant modulus, and this method also has to be adapted to work with pseudo-residues.  We describe the resulting algorithm in \Sec{LS-mmrns}.
\subsection{\label{LSSgbe}Generalized base extension with a redundant modulus}
Base extension refers to the operation of computing a residue of a number modulo a new modulus from a given RNS representation. 
We need a slight generalization of the base extension method from \cite{sk-fbe}.
Our generalization is based on the following.
\bpn{LPbasic}\rm
Let $\cB=(M_1, \ldots, M_k)$ be an RNS, with dynamical range $M=M_1\cdots M_k$, let $\cI=[-1+e, e)$ be a residue interval,
and let $\phi$ be an expansion constant. Let $x$ be an integer with $x\in M\cI$, and let $\ga_1, \ldots, \ga_k$ be pseudo-residues with $x\equiv \ga_i(M/M_i)$ and $\ga_i\in \phi \cI$ for all $i$. 
Then $x$ can be written as 
\beql{LE-crt} x=\sum_{i=1}^k \ga_i (M/M_i) -qM\eeql
for an integer $q$ with $-k(1-e) \phi -e<q<k e\phi +1-e$.
\epn
\bpf
The CRT states that if $\ga_1, \ldots, \ga_k$ are integers for which $\ga_i(M/M_i)\equiv x\bmod M_i$ for $i=1, \ldots, k$, then $x$ is of the form
(\ref{LE-crt})
with $q$  integer.
Write $\xhat=\sum_{i=1}^k r_i(M/M_i)$. Now $\ga_i(M/M_i)\in \phi M\cI$, hence
$\xhat\in k\phi M\cI$, and $x=\xhat-q M\in M\cI$ by assumption. Hence $q=\xhat/M - x/M$ satisfies $(-1+e) k\phi -e < q <k\phi e +1-e$.
\epf
\bco{LCbasic}\rm In \Pn{LPbasic}, we have  $e=1$ and $q\in k\phi \cI$ in the case of standard residues, and $e=1/2$ and $q\in (k\phi+1) \cI$ in the case of symmetric residues. Hence 
\beql{LEbasic} q=\red{(\red{-M^{-1}}{M_0})\red{x}{M_0} +\sum_{i=1}^k \red{r_i}{M_0}(\red{M_i^{-1}}{M_0})}{M_0}\eeql
if $M_0\geq \lceil k\phi\rceil$ (standard residues) or $M_0\geq \lceil k\phi\rceil+1$ (symmetric residues). 
\eco
As a consequence of \Pn{LPbasic} and \Co{LCbasic}, by combining (\ref{LE-crt}) and (\ref{LEbasic}) we can use the pseudo-residues $\ga_1, \ldots, \ga_k$ of an integer $x\in \cI M$  together with the  residue $x_0=\red{x}{M_0}$ for a ``redundant'' modulus $M_0$, with $M_0\geq   \lceil k\phi\rceil$ (standard residues) or $M_0\geq \lceil k\phi\rceil+1$ (symmetric residues), to find a (pseudo-)residue $\rho$ for a new modulus $n$, by using that
\[\rho\equiv \sum_{i=1}^k \ga_i(M/M_i)-qM \bmod n.\]

\subsection{The Bajard-Imbert  Montgomery RNS algorithm}
Our method is based on an algorithm similar to the  RNS-based Montgomery multiplication algorithm described in \cite{bi-rns}, referred to here as the {\em Bajard-Imbert RNS algorithm\/}.
This algorithm employs an RNS consisting of a {\em left RNS\/} $\cB=(M_1, \ldots, M_k)$ with dynamical range $M=M_1M_2\cdots M_k$, a {\em right RNS\/} $\cB'=(M_{k+1}, \ldots, M_{2k})$ with dynamical range $M'=M_{k+1}M_{k+2}\cdots M_{2k}$, and a {\em redundant modulus\/} $M_0$ used for base extension as in \cite{sk-fbe},
and computes a Montgomery multiplication with Montgomery constant $M$, for a modulus~$N$ that satisfies the conditions $0<(k+2)^2N<\min(M,M')$ and $\gcd(N,M)=1$.
Given 
inputs $x,y$ represented by their residues in $\{M_0\}\cup\cB\cup\cB'$, a representation of a Montgomery product $z\equiv xyM^{-1} \bmod N$ in $\{M_0\}\cup\cB\cup\cB'$ is computed using the following steps.
\ben
\item
Compute $h=xy$ in $\{M_0\}\cup\cB\cup\cB'$.
\item
Compute $\mu_i=\red{-N^{-1}(M/M_i)^{-1}h}{M_i}$ for $i=1, \ldots, k$; set 
$u=\sum_{i=1}^k \mu_i(M/M_i)$.
Then $h+uN\equiv 0\bmod M$.
\item
Compute the representation of $u$ in $\{M_0\}\cup\cB'$.
\item
Compute $z=(h+uN)/M$ in $\{M_0\}\cup\cB'$.
\item
Find the representation of $z$ in $\cB$ (by base extension).
\een
The algorithm from  \cite{bi-rns} has the property that if the inputs $x,y$ satisfy $0\leq x,y<(k+2)N$, then the result~$z$ of the Montgomery multiplication again satisfies $0\leq z<(k+2)N$. 


Note that in order to apply this algorithm for large (2048-bits) moduli $N$, we require that $M/(k+2)^2$ is large, which is not possible by employing small moduli $M_i$ only. To enable efficient implementation of the required modular arithmetic, one solution would be to choose moduli of a simple form such as $2^n-c$ with small~$c$ or as $2^n\pm 2^m\pm1$ \cite{bdem-rns}. 
In this paper, we propose to implement the arithmetic modulo the $M_s$ by using a similar RNS-based algorithm,  now employing an RNS with much smaller moduli. Since Montgomery multiplication 
does not deliver exact residues, the RNS algorithm that we use  should now be able to handle 
RNS representations made up from such ``pseudo-residues''. 
\section{\label{LS-mmrns}The new algorithm}
\subsection{\label{LSS-alg1}The recursive RNS method - The basic assumptions}
Our method builds an RNS implementation of modular arithmetic for new (larger) moduli on top of a layer of (smaller) RNS moduli for which some form of modular arithmetic has already been realized. On the lowest (bottom) level, we assume that all moduli have size at most $2^t$ and that all the modular arithmetic is done by lookup tables of size $2^t\times 2^t$. Here, allowing entries with a slightly wider range permits to use residues modulo one modulus as entries to a table for another modulus, which our algorithm requires. 


We now state exactly what we require when constructing a new RNS layer. Let $\cI=[-e, 1-e)$ be an interval of length 1, let $m$ be a positive integer, and let $B_1, \gvf_1, \phi_1$ be positive rationals, with $\gvf\geq \phi_1\geq1$.
We will write  $\cA(B_1,m,\cI,\gvf_1,\phi_1)$ to denote  that for all moduli~$n$ with $1\leq n\leq B_1$ and $\gcd(n,m)=1$,  
the following statements hold.

\vspace{.02in}
%
\noindent
1.\ ({\em Montgomery product)\/}
For all integers $x,y\in \gvf_1 n\cI$, we can compute an integer
$z=x\otimes_{(n,m)} y$
for which $z\equiv xym^{-1} \bmod n$ and $z\in \gvf_1 n\cI$; moreover, if $y\in n\cI$, then even $z\in \phi_1 n\cI$.

\vspace{.02in}
\noindent
2.\ ({\em Modular multiply-and-accumulate\/}) 
Given integer constants $c^{(1)}, \ldots, c^{(k)}\in n\cI$ for all $i$, then for all $k$-tuples of integers $x^{(1)}, \ldots, x^{(k)}\in \gvf n\cI$, we can compute 
an integer
$\xi=\cS(c^{(1)}, \ldots, c^{(k)}; x^{(1)},\ldots , x^{(k)})$
for which
$\xi\equiv c^{(1)}x^{(1)}+\cdots +c^{(k)}x^{(k)}\bmod n$ and $\xi\in \gvf_1 n\cI$.

\vspace{.02in}

\noindent
Note that the above assumptions hold when we construct the {\em bottom\/} level RNS, with $B_1=2^t$, $m=1$, and $\gvf_1=\phi_1=1$. When constructing higher levels, both Montgomery multiplication and modular multiply-and-accumulate can be realized with the aid of {\em Montgomery reduction\/}, that is, if the following asumption holds.

\vspace{.02in}

\noindent
3. ({\em Montgomery reduction\/})
For all integers $h\in  \gvf_1^2 n^2\cI$, we can compute an integer
$r=\cR_{(n,m)}(h)$
for which $r\equiv hm^{-1} \bmod n$ and $r\in \gvf_1 n\cI$; moreover, if $h\in \gvf n^2\cI$, then even $r\in  \phi_1 n\cI$. 

Note that the somewhat strange-looking assumption on~$h$ enables us to define $x\otimes_{(n,m)} y=\cR_{(n,m)}(xy)$, provided that we can compute the product $h=xy$. Moreover, Montgomery reduction can be used to scale down intermediate computational results, where the extra modular factor $m^{-1}$ incurred by the reduction can either be incorperated in the representation constants or be compensated for by modifying the constants $c^{(i)}$. 
We now sketch how to do the latter, leaving some details to the reader.  If $k\leq \gvf_1$, then put $d^{(i)}=\red{mc^{(i)}}{n}$ for all~$i$; then the integer $h=d^{(1)}x^{(1)}+\cdots +d^{(k)}x^{(k)}$ satisfies $h\in k\gvf_1 n^2\cI^2\subseteq \gvf_1^2n^2\cI^2$, so we can take $\xi=\cR_{(n,m)}(h)$. If not, then put $d^{(i)}=\red{m^2c^{(i)}}{n}$ for all $i$, and partition the index set $\{1, \ldots, k\}$ into sets~$I_j$ for which $|I_j|\leq \gvf$, so that $h^{(j)}=\sum_{i\in I_j}d^{(i)} x^{(i)}\in \gvf^2n^2\cI^2$; now compute $\xi^{(j)}=\cR_{(n,m)}(h^{(j)})$ for all $j$, set $h=\sum_j \xi^{(j)}$, and let $\xi=\cR_{(n,m)}(h)$. Again, this works provided that we can guarantee that $h\in\gvf^2n^2\cI^2$.  If $h$ can still be too big, then we do the reduction in even more stages, including even more factors $m$ in the constants $d^{(i)}$ to compensate for the Montgomery reductions.

From the above, we see that if assumption $\cA(B_1,m,\cI, \gvf_1,\phi_1)$ holds, then we can realize modular arithmetic with expansion bound $\gvf_1$ and \resi interval~$\cI$. Strictly speaking, it is not necessary to include the constant $\phi_1$ in the assumption. Indeed, note that we could simply {\em define\/} $\phi_1$ as the smallest integer for which it is true that $\cR_{(n,m)}(h)\in\phi_1 n\cI$ whenever $h\in \gvf_1 n^2\cI^2$,  for every integer $n$ with  $1\leq n\leq B_1$ and $\gcd(n,m)=1$. 
\subsection{\label{LSS-alg2}The recursive RNS method - choice of RNS moduli}
Suppose that the assumption $\cA(B_1,m,\cI,\gvf_1,\phi_1)$ in \Sec{LSS-alg1} above 
holds.
We aim to show that we can satisfy such assumptions 
for some new Montgomery constant $M$ and  some $B\gg B_1$, and some new expansion constants $\gvf, \phi$, by adding a new RNS layer on top of the existing layers.
To this end, 
we first choose a {\em left RNS\/} $\cB=(M_1, \ldots, M_k)$ with dynamical range $M=M_1 \cdots M_k$,  a {\em right RNS\/} $\cB'=(M_{k+1}, \ldots, M_{k+l})$ with dynamical range $M'=M_{k+1} \cdots M_{k+l}$, and a redundant modulus $M_0$. 
The moduli have to be chosen such that the following is satisfied.
\bit
\item[$\bullet$] $M_s\leq B_1$ and $\gcd(M_s,m)=1$ for $s=1, \ldots, k+l$, so the modular arithmetic modulo every $M_s$ can be realized;
\item[$\bullet$]
The full base $\cB^*=\{M_0\}\cup \cB\cup \cB'$ is an RNS, that is, $\gcd(M_s, M_t)=1$ for all $0\leq s<t\leq k+l$;
\item[$\bullet$]
The arithmetic modulo the redundant modulus $M_0$ is {\em exact\/}, that is, every computed  residue modulo $M_0$ is contained in 
an interval of size $M_0$. 
\eit
For example, the modulus $M_0$ can be ``small'', so that the arithmetic modulo $M_0$ can be done by table lookup, or $M_0$ can be the product of several ``small'' moduli. Another possibility is to take $M_0$ of a simple form, for example of the form $2^r-1$ or even $2^r$. Additional constraints on the redundant modulus will be discussed later.

Let $D=M_0MM'$ denote the dynamical range of the RNS~$\cB^*$. 
In what follows, given constants $H_0=1, H_1, \ldots, H_{k+l}$ with $\gcd(H_s,M_s)=1$ for all $s\geq1$, we refer to a $H$-representation $(\ga_0, \ga_1, \ldots, \ga_{k+l})$ in $\cB^*$ for an integer $x$ with $x\in D\cI$ as a {\em $(H,\gvf_1)$-representation\/} if the pseudo-residues $\ga_s$ used in the representation satisfy $\ga_0=\red{x}{M_0}$ (so $\ga_0$ is a true residue modulo $M_0$) and $x\equiv \ga_sH_s\bmod M_s$ with $\ga_s\in\gvf_1 M_s \cI$ for $s=1, \ldots, k+l$.
\subsection{\label{LSS-alg3}Alternative RNS representations}
When introducing Montgomery multiplication, we have already discussed the special representation of a residue or pseudo-residue $X$ modulo $N$ by its Montgomery representation $x=RX$, where $R$ is the Montgomery constant. To represent the 
(pseudo-)residues of the inputs and outputs  with respect to the RNS $\cB^*=\{M_0\}\cup \cB\cup\cB'$, we generalize this idea. Let $D=M_0MM'$ denote the dynamical range of $\cB^*$. 
\bde{LDpropH} \rm Given constants $H_0=1, H_1, \ldots, H_{k+l}$ with $\gcd(H_s,M_s)=1$ for $s=1, \ldots, k+l$, we define a {\em $(H,\cI, \gvf_1)$-representation\/} of an integer $x$  to be a representation $(\ga_0; \ga_1, \ldots, \ga_{k+l})$ of~$x$ with residue $\ga_0=\red{v}{M_0}\in M_0\cI$ and pseudo-residues $\ga_s\in \gvf_1M_s\cI$ for which $x\equiv H_s \ga_s \bmod M_s$, for $s=1, \ldots, k+l$.  
\ede
If the \resi interval $\cI$ 
is clear from the context, 
then we simply refer to such a representation as a {\em $(H,\gvf_1)$-representation\/}.
Note that a proper $H$-representation of~$x$ indeed uniquely determines $x$ if we know beforehand that $x\in D\cI$.

An obvious choice would be to represent all the residues in Montgomery representation with respect to the Montgomery constant $m$, that is, to take $H_s=K_s=m^{-1}\bmod M_s$ for $s=1, \ldots, m$. In that case, we can compute the Montgomery representation $\chi_s$ of a residue modulo $M_s$ of the ordinary product $h=xy$  directly from the Montgomery representations $\ga_s, \gb_s$ of the residues modulo $M_s$ of $x$ and~$y$ as $\chi_s=\ga_s\otimes_{(M_s,m)} \gb_s$, which is assumed to be an available operation.  
This choice helps to avoid certain scaling operations in the algorithms. We will see later that there is another, less obvious choice for the representation constants $H_s$  that can significantly lower the computational complexity of the algorithm. 
\subsection{\label{LSS-alg4}The recursive RNS method - the method}
%
We will now describe our method to implement Montgomery reduction and Montgomery multiplication modulo $N$ for certain moduli $N$,
given that the assumption $\cA(B_1,m,\cI,\gvf_1,\phi_1)$ holds. 
A high-level description of the algorithm to compute $z=x\otimes_{(N,M)}y$ consists of the following steps.
\ben
\item
Compute $h=xy$ by computing the residue of $h$ modulo $M_0$ and suitable pseudo-residues of $h$ in $\cB\cup \cB'$;
\item
compute $z=\cR_{(N,M)}(h)$ as follows:
\ben
\item
compute $u$ such that $h+uN\equiv 0\bmod M$ by computing suitable pseudo-residues in the left RNS~$\cB$;
\item
compute $z=(h+uN)/M$ by computing the residue modulo $M_0$ and suitable pseudo-residues in the right RNS $\cB'$;
\item
use base extension to compute corresponding pseudo-residues of $z$ in the left RNS $\cB$.
\een
\een
Below we work out these steps in detail. In this section, we concentrate on explaining and verifying that the computed RNS representations indeed satisfy the required congruences. As a consequence, the algorithm works provided that the numbers thus represented coincide with the numbers that they are supposed to represent, that is, provided that the numbers that we want to compute are known beforehand to be  contained in the ``correct''  interval. The required bounds that guarantee this are analyzed in~\Sec{LS-B}.

Let $D=M_0MM'$ denote the dynamical range of the full  RNS $\cB^*=\{M_0\}\cup \cB\cup \cB'$.
%
Fix representation constants $H_0=1, H_1, \ldots, H_{k+l}$ and $K_0=1, K_1, \ldots, K_{k+l}$ with $\gcd(H_s, M_s)=\gcd(K_s,M_s)=1$ for all $s$.   
In addition,  we choose (small) integers  $S_1, \ldots, S_k$ (the reason for this will be discussed later; for the time being, we may assume that $S_i=1$ for all $i$). 
Suppose we are given  inputs $x,y\in\gvf N\cI$  with $(H,\gvf_1)$-representations $(\ga_0,\ldots, \ga_{k+l})$ and $(\gb_0, \ldots, \gb_{k+l})$, respectively. Then we proceed as follows.

\vspace{.05in}
\noindent
1. Compute $h_0=\chi_0=\red{\ga_0\gb_0}{M_0}$ and 
\beql{LE-chis}\chi_s=\ga_s\otimes_{(M_s,m)}\gb_s\eeql
for $s=1, \ldots, k+l$.
Then $h\equiv xy \equiv \ga_s\gb_sH_s^2  =\chi_s m H_s^2\bmod M_s$, that is, $(\chi_0, \chi_1, \ldots, \chi_{k+l})$ is a $(K,\gvf_1)$-representation, 
where 
$K_0=1$ and $K_s=mH_s^2$ for $s=1, \ldots, k+l$.

\vspace{.05in}
\noindent
2. Given a proper $K$-representation $(\chi_0, \chi_1, \ldots, \chi_{k+l})$ for $h$, for certain representation constants $K_s$, 
we compute $z=\cR_{(N,M)}(h)$ as follows.
\ben 
\item
Compute 
\beql{LE-mui}\mu_i=\chi_i\otimes_{(M_i,m)} \red{-N^{-1}K_i(M/M_i)^{-1}S_i^{-1}m}{M_i}\eeql
for $i=1, \ldots, k$. Then $\mu_iS_i\equiv -N^{-1}h(M/M_i)^{-1}\bmod M_i$ for $i=1, \ldots, k$, so by the CRT, 
the integer
\beql{LE-u} u=\sum_{i=1}^k \mu_iS_iM/M_i\eeql
satisfies $u\equiv -N^{-1}h\bmod M$, that is,   $h+uN\equiv 0\bmod M$.
%
\item
Next, let $z=(h+uN)/M\equiv hM^{-1}+uNM^{-1}\bmod M'$. We want to compute a $(H,\gvf_1)$-representation $(\xi_0, \xi_1, \ldots, \xi_{k+l})$ for $z$.
To this end,  compute 
\beql{LE-xi0}\xi_0=\red{\chi_0 \red{M^{-1}}{M_0}+\sum_{i=1}^k \mu_iS_i \red{N M_i^{-1}}{M_0}}{M_0}\eeql
and determine $\xi_j$ for $j=k+1, \ldots, k+l$ such that
$\xi_j\equiv zH_j^{-1}  \equiv \chi_jK_jM^{-1}H_j^{-1}+\sum_{i=1}^k \mu_iS_iN M_i^{-1} H_j^{-1} \bmod M_j$, 
by computing
\begin{multline} \label{LE-sj} \xi_j \equiv  \chi_j\red{K_jM^{-1}H_j^{-1}}{M_j}\\
+\sum_{i=1}^k \mu_i\red{S_iN M_i^{-1} H_j^{-1} }{M_j} \bmod M_j
\end{multline}
with $\xi_j\in \gvf_1 M_j\cI$, using the multiply-and-add operation $\cS_{M_j}$ from Section~\ref{LSS-alg1}.
\item
We have now determined the part of the $H$-representation of $z$ for the right RNS. 
To find the part of the $H$-representation of $z$ for the left RNS, we use (generalized) base-extension as discussed in Section~\ref{LSSgbe}.
\bit
\item
First, we write $z$ in the form
\beql{LE-zeta} z=\sum_{j=k+1}^{k+l} \eta_j (M'/M_j) -qM'.\eeql
By the 
constructive CRT,
we should take $\eta_j\equiv z(M'/M_j)^{-1} \equiv \xi_j H_j (M'/M_j)^{-1} \bmod M_j$ for $j=k+1, \ldots, k+l$. 
Hence we should take
\beql{LE-etaj} \eta_j=\xi_j \otimes_{(M_j,m)} \red{H_j(M'/M_j)^{-1} m}{M_j}\eeql
for $j=k+1, \ldots, k+l$.
%
%
\item
Then, we use the redundant residue $\xi_0=z_0$ to determine $q$ from 
\beql{LE-q}
q\equiv \xi_0\red{(-M')^{-1}}{M_0}+ \sum_{j=k+1}^{k+l} \eta_j \red{M_j^{-1}}{M_0}\bmod M_0.
\eeql
\item
And finally, we use (\ref{LE-zeta}) to compute
\begin{multline} \xi_i\equiv zH_i^{-1}\equiv \red{-M' H_i^{-1}}{M_i}\\
+ \sum_{j=k+1}^{k+l} \eta_j \red{(M'/M_j)H_i^{-1}}{M_i}\bmod M_i, 
\end{multline}\label{LE-ti}
with $\xi_i\in \gvf_1 M_i\cI$, again by using the multiply-and-add operation $\cS_{M_i}$ from Section~\ref{LSS-alg1}.
\eit
\een
Preferably, on a non-bottom level we compute (\ref{LE-sj}) by computing
\beq s_j=\chi_j\red{K_jM^{-1}H_j^{-1}m}{M_j}
+\sum_{i=1}^k \mu_i\red{S_iN M_i^{-1} H_j^{-1}m}{M_j}
\eeq
followed by
\beq \xi_j=\cR_{(M_j,m)}(s_j)\eeq
and  
(\ref{LE-ti}) by computing
\beq t_i= q\red{-M' H_i^{-1}m}{M_i}+ \sum_{j=k+1}^{k+l} \eta_j \red{(M'/M_j)H_i^{-1}m}{M_i} \eeq
followed by 
\beq\xi_i=\cR_{(M_i,m)}(t_i).\eeq

We will refer to this method to compute the $\xi_j$'s 
and the $\xi_i$'s as {\em postponed reduction\/}. 
This is similar to the method called {\em accumulate-then-reduce\/} \cite{lim2000fast}, also called {\em lazy reduction\/} (see, e.g., \cite{aranha2011faster}), for computing a sum-of-products where modular reduction is done only once at the end, instead of after each multiplicaton and addition. We will discuss postponed reduction in more detail in \Sec{LSpp}.
%
%
\subsection{\label{LSS-alg4}The recursive RNS method - the algorithm}
The description above can be summarized as follows.
Assume that we are given  inputs $x,y\in \gvf N\cI$ by means of $(H,\gvf_1)$-representations $(\ga_0,\ldots, \ga_{k+l})$ and $(\gb_0, \ldots, \gb_{k+l})$, respectively.  In order to compute $h=xy$, we run Algorithm~1 below. 

\begin{algorithm}[ht]
\caption{Computation of $h=xy$}
\label{algo1}
\begin{algorithmic}[1]
\algloop{For}
\State $\chi_0=\red{\ga_0\gb_0}{M_0}$ \ 
\For{$s=1$ \To $k+l$} \  $\chi_s=\ga_s\otimes_{(M_s,m)}\gb_s$
\end{algorithmic}
\end{algorithm}

\noindent
This has the following result.
\btm{LTh}\rm 
The tuple $(\chi_0, \chi_1, \ldots, \chi_{k+l})$ constitutes a $(K,\gvf_1)$-representation for $h=xy$ in $\cB^*$ provided that $h\in M_0MM' \cI$,
where $K_0=1$ and $K_s=mH_s^2$ for $s=1, \ldots, k+l$.
\etm
\bpf 
We have $\chi_0\equiv \ga_0\gb_0\equiv xy= h\bmod M_0$ and 
$\chi_s=\ga_s\otimes_{(M_s,m)}\gb_s\equiv H_s^{-1}x H_s^{-1}ym^{-1}\equiv K_s^{-1}h\bmod M_s$ for $s=1, \ldots, k+l$.
\epf
Next, assume that $h$ has $(K,\gvf_1)$-representation $(\chi_0, \chi_1, \ldots, \chi_{k+l})$. We desire to compute a $(H,\gvf_1)$-representation $(\xi_0, \ldots \xi_{k+l})$ for $z=\cR_{(N,M)}(h)$ as above, with $z\in \gvf N\cI$ again. To achieve that, proceed as follows.  First, choose $S_1, \ldots, S_k$ with $S_i$ small for all $i$ (the reason will be discussed later). 
Next, pre-compute the constants
\bit
\item[$\bullet$]
$C_i=\red{-N^{-1}K_i(M/M_i)^{-1}S_i^{-1}m}{M_i}$  \ \ $(i=1, \ldots, k)$;
\item[$\bullet$]
$D_{0,0}=\red{M^{-1}}{M_0}$ and\\
$D_{0,i}=\red{S_iM_i^{-1}N}{M_0}\quad(i=1, \ldots, k)$,
\item[$\bullet$]
for  $j=k+1, \ldots, k+l$, \\
$D_{j,0}=\red{K_{j}M^{-1}H_j^{-1}}{M_{j}}$ and\\
$D_{j,i}=\red{S_iM_i^{-1}NH_{j}^{-1}}{M_{j}}\quad(i=1, \ldots, k)$ 
\item[$\bullet$]
$E_j=\red{H_{j}(M'/M_j)^{-1}m}{M_{j}} \qquad (j=k+1, \ldots, k+l)$;
\item[$\bullet$]
$F_{0,0}=\red{(-M')^{-1}}{M_0}$ and\\ 
$F_{0,j}=\red{M_{j}^{-1}}{M_0} \quad (j=k+1, \ldots, k+l)$;
\item[$\bullet$]
for $i=1, \ldots, k$, \\
$G_{i,0}=\red{-M'H_i^{-1}}{M_i}$ and\\
$G_{i,j}=\red{(M'/M_j)H_{i}^{-1}}{M_i} \quad (j=k+1, \ldots, k+l) $,
\eit
%
%
and then, run Algorithm 2 below, using the modular add-and-accumulate operator $\cS$ discussed in \Sec{LSS-alg1}.
\begin{algorithm}[ht]
\caption{Computation of $z=\cR_{(N,M)}(h)$}
\label{algo2}
\begin{algorithmic}[1]
\For{$i=1$ \To $k$} 
\State $\mu_i=\chi_i\otimes_{(M_i,m)}C_i$
\EndFor
\State $\xi_0=\red{\chi_0D_{0,0}+\mu_1D_{0,1}+\cdots+\mu_kD_{0,k}}{M_0}$
\For{$j=k+1$ \To $k+l$}
\State $\xi_{j}=\cS(D_{j,0}, D_{j,1}, \ldots, D_{j,k}; \chi_{j}, \mu_1, \ldots, \mu_k)$
\State $\ \ \ \, \equiv \chi_{j}D_{j,0}+\mu_1D_{j,1}+\cdots +\mu_k D_{j,k}\bmod M_j$
\EndFor
\For{$j=k+1$ \To $k+l$}
\State $\eta_{j}=\xi_{j}\otimes_{(M_{j},m)}E_{j}$
\EndFor
\State $\eta_0=\red{\xi_0F_{0,0}+\mu_{k+1}F_{0,k+1}+\cdots +\mu_{k+l}F_{0,k+l}}{M_0}$
\For{$i=1$ \To $k$}
\State $\xi_i=\cS(G_{i,0}, G_{i,k+1}, \ldots, G_{i,k+l}; \eta_0, \eta_{k+1}, \ldots, \eta_{k+l})$
\State $\ \ \ \, \equiv\eta_0G_{i,0}+\eta_{k+1}G_{i,k+1}+\cdots+\eta_{k+l}G_{i,k+l} \bmod M_i$
\EndFor
\end{algorithmic}
\end{algorithm}

It is not difficult to verify that by this choice of  constants, 
the output $(\xi_0,\ldots ,\xi_{k+l})$ of Algorithm 2 has the following properties.
\btm{LTz}\rm Define $u=\sum_{i=1}^k\mu_iS_i(M/M_i)$ and $z=(h+uN)/M$. Then $u\equiv -N^{-1}h \bmod M$ and hence $z$ is integer. We have  $\xi_0=\red{z}{M_0}$ and $z\equiv \xi_j H_j\bmod M_j$ for $j=k+1, \ldots, k+l$,
and hence 
$z=\sum_{j=k+1}^{k+l} \eta_j(M'/M_j)-qM'$ for some integer $q$. We have  $q\equiv \eta_0\bmod M_0$, and, setting $z'=\sum_{j=k+1}^{k+l} \eta_j(M'/M_j)-\eta_0M'=z+(q-\eta_0)M'$, we have that  $z'\equiv \xi_iH_i\bmod M_i$ for $i=1, \ldots, k$. 
\etm
\bpf
Straightforward from the definitions of the $\mu_i$ and the $\xi_s$, and from the definitions of the various constants.  
We have that $\mu_i\equiv \chi_i\otimes_{(M_i,m)}C_i  \equiv   hK_i^{-1}  (-N^{-1})K_i(M/M_i)^{-1}S_i^{-1}m m^{-1}\equiv -hN^{-1}(M/M_i)^{-1}S_i^{-1} \bmod M_i$, so that $u\equiv \mu_iS_i(M/M_i)\equiv -N^{-1}h\bmod M_i$. Hence $u\equiv -hN\bmod M$,  so that $z$ is integer.

We have  $z=hM^{-1}+uNM^{-1}=hM^{-1}+\sum_{i=1}^k \mu_i S_iM_i^{-1}N$, hence 
$z\equiv \chi_0D_{0,0}+\sum_{i=1}^k \mu_iD_{0,i} \equiv \xi_0\bmod M_0$ and 
 $z\equiv \chi_jD_{j,0} H_j+\sum_{i=1}^k\mu_i D_{j,i} H_j \equiv \xi_jH_j\bmod M_j$ for $j=k+1, \ldots, k+l$.

Next,  since $\eta_j=\xi_{j}\otimes_{(M_{j},m)}E_j\equiv m^{-1}zH_j^{-1}H_j(M'/M_j)^{-1}m=z(M'/M_j)^{-1}$,
we have $z'\equiv \eta_j (M'/M_j)\equiv z\bmod M_j$. Also, $\eta_0M'\equiv -\xi_0+\sum_{j=k+1}^{k+l}\eta_jMM_j^{-1}\equiv  -z+(z'+\eta_0M')\bmod M_0$, hence $z\equiv z'\bmod M_0$ and $q\equiv \eta_0 \bmod M_0$.

Finally, for $i=1, \ldots, k$, we have $z'H_i^{-1}m\equiv \mu_0(-M'H_i^{-1}m)+\sum_{j=k+1}^{k+l} \mu_j (M'/M_j) H_i^{-1}m \equiv t_i\bmod M_i$.
\epf

Note that the Algorithm 2 can be implemented with one register of length $1+k+l$ to store the values $\eta_0$, $\mu_i$ for $i=1, \ldots, k$, 
and $\eta_j$ for $j=k+1, \ldots, k+l$, 
and another register of length $1+k+l$ to store the values of $\chi_s$ for $s=0, 1, \ldots, k+l$, which can be overwritten to also store the $\xi_s$ for $s=0, 1, \ldots, k+l$.

Note also that in order to change the modulus $N$, we can simply replace some of the constants in algorithm 2. 
\subsection{The redundant modulus}
Note that to be able to execute Algorithm 2, the modulus~$M_0$ has to have some additional properties.
\ben
\item In steps 4 and 12, we need to be able to extract the residue modulo $M_0$ from the numbers $\mu_1, \ldots, \mu_k$; so either these numbers are small, or this residue must be obtainable from one or more of the residues in an RNS representation for these numbers. 
\item In step 15, we have to be able to multiply a constant modulo $M_i$ with a computed residue $\eta_0$ modulo $M_0$. 
\een
These requirements are indeed satisfied when adding the first (bottom) RNS layer by our assumptions at the start of \Sec{LSS-alg1}. Suppose that on the bottom level, we have a redundant modulus $m_0$, a left RNS with moduli $m_1, \ldots, m_{k_1}$, and a right RNS with moduli $m_{k_1+l_1}, \ldots, m_{k_1+l_1}$. Let $m=m_1\cdots m_{k_1}$ denote the dynamical range of the left RNS (this will be the Montgomery constant for the moduli on the next level). For the redundant modulus $M_0$ on the second level, we can take for example $M_0=m_0m_j$ for some $j>0$.  Since every pseudo-residue $\mu_i$ on the second level is represented by its residues in the full RNS on the bottom level, we can immediately obtain the residues of $\mu_i$ modulo $m_0$ and modulo $m_j$, and  by the CRT, these two residues represent the residue modulo $M_0$. Similarly, a constant $C$ modulo some $M_i$ (that is, an integer $C\in M_i \cI$) is represented by its residues $C_s=\red{C}{m_s}$ modulo the $m_s$.  To multiply by a residue $\eta_o$, use Mixed Radix Conversion (see, e.g., \cite{st-ra}) to write $\eta_0$ in the Mixed Radix form $a+m_j b$. Then $\eta_0C$ has residues $\red{(a+m_j b) C_s}{M_s}$, where each residue can be obtained by four modular operations.

On higher levels, Step 2 above can always be executed in a similar way by obtaining some kind of Mixed Radix representation for $\eta_0$, as long as there is no overflow on any level. And to be able to execute Step 1 above, we should probably require that on higher levels the redundant modulus is the product of some of the moduli on the level below. Further implementation considerations are left to the reader.


\section{\label{LS-B}Bounds}
For the algorithms to work as desired, several bounds have to hold. We need some preparation. Let $\cI=[-1+e, e)$ be a \resi\ interval, with $e=1$ (standard residues) or $e=1/2$ (symmetric residues). Note that $\cI^2=e\cI$ if $e=1/2$ or $e=1$.
We also need a bound on the integer $u=\sum_{i=1}^k \mu_i S_i (M/M_i)$ as defined in \Tm{LTz}. We let $U$ denote the smallest positive integer for which $u\in UM\cI$. We have to ensure that $U$ exists. In the case of standard residues, this is achieved by requiring that the numbers $S_i$ are all positive, with $0<S_i\leq S$, for some number $S$;  in the case of symmetric residues, we assume that $|S_i|\leq S$, with at least one $S_i$ equal to 1. Since $\mu_i\in \phi_1 M_i\cI$, in both  cases we can take $U=\phi_1 k S$. We are now ready to state our main result. 
\btm{LTbounds}\rm
 Given the above notation, put $\gvf=U/\gre$ with $0<\gre<1$, and let $\phi=U+1-\gre$. Suppose that $N\leq M\gre(1-\gre)e^{-1}/U$, $M'\geq M(1-\gre)e^{-1}$, and $M_0\geq \lceil l \phi_1 \rceil$ ($e=1$, standard residues) or $M_0\geq 1+\lceil l \phi_1 \rceil$ ($e=1/2$, symmetric residues). Then given $(H,\gvf_1)$-representations for $x,y\in \gvf N \cI$, Algorithm 1 produces a $(K,\gvf_1)$-representation for $h=xy\in\gvf^2N^2 \cI\subseteq M_0MM'\cI$ and 
given a $(K,\gvf_1)$-representation for $h\in \gvf^2 N^2\cI^2$,  Algorithm 2 produces a $(H, \gvf_1)$-representation $(\xi_0, \ldots, \xi_{k+l})$  for~$\cR_{(N,M)}(h)=z$ with $z\in  \gvf N\cI$; moreover, even $z\in \phi N\cI$ if $h\in \gvf N^2\cI^2$.
\etm
\bpf
Suppose that $h\in \gvf^2N^2\cI^2$. We have $z=(h+uN)/M$ with $u=\sum_{i=1}^k\mu_iS_i(M/M_i)\in UM\cI$ by our above assumptions.  
Hence $z\in ( \gvf^2N^2e+ U MN)/M\cI$.
So we have that $z\in  \gvf N\cI$ provided that $\gvf^2N/M +U\leq \gvf$. From this inequality, we see that $\gvf>U$.  So we can write $\gvf=U/\gre$ with $0< \gre <1$,  and the condition becomes 
\beql{LEnc} N\leq M\gre(1-\gre)e^{-1}/U=M(1-\gre)e^{-1}/\gvf.\eeql
If (\ref{LEnc}) holds, then $\gvf N<2M$ and $\gvf N\leq  M'$, hence $\gvf^2 N^2<2MM'\leq M_0MM'$, hence $xy\in\gvf^2N^2\cI^2\subseteq M_0MM'\cI$. 
So  according to \Tm{LTh}, $(\chi_0, \chi_1, \ldots, \chi_{k+l})$ is a $(K,\gvf_1)$-representation for $h=xy$ in~$\cB^*$. Using (\ref{LEnc}), it is easily checked that even $z\in \phi N\cI$ if $h\in \gvf N^2\cI^2$.

Finally, since $z\in \gvf N \cI\subseteq M' \cI$, the bound on $M_0$ follows from \Pn{LPbasic} and \Co{LCbasic}.
%
\epf
As a consequence of \Tm{LTbounds}, we can can again satisfy assumption $\cA(B,m,\cI,\gvf,\phi)$ (see \Sec{LSS-alg1}), now with $B=M\gre(1-\gre)e^{-1}/U\gg B_1$, Montgomery constant $M$, and expansion constants $\gvf=U/\gre$ and  $\phi=U+1-\gre=\gvf \gre +1-\gre \leq \gvf$ since $\gvf\geq1$.
%
\section{Improvements}
In this section, we discuss several ways in which the algorithm can be optimized or improved.
\subsection{\label{LSpp}Postponed reduction}
Under certain conditions, steps 6 and 14 in Algorithm~2 can be done by Montgomery reduction. For example, on a non-bottom layer, step 6 may be replaced by\\
$t_j=  \chi_{j}D'_{j,0}+\mu_1D'_{j,1}+\cdots +\mu_k D'_{j,k}$;\\
$\xi_j=\cR_{(M_j,m)}(t_j)$,\\
where $D'_{j,i}=\red{mD_{j,i}}{M_j}$ for all $i$. 
This  will work provided that $t_j$ can be computed and satisfies $t_j\in \gvf_1^2M_j^2\cI^2$.
This is similar to the method called {\em accumulate-then-reduce\/} \cite{lim2000fast}, also called {\em lazy reduction\/} (see, e.g., \cite{aranha2011faster}), for computing a sum-of-products where modular reduction is done only once at the end, instead of after each multiplicaton and addition. 
This is a special case (in fact the simplest case) of the possible implementation of the multiply-and-accumulate operation $\cS$ as discussed in \Sec{LSS-alg1}. Since $\chi_j\in \gvf_1 M_j \cI$, $D'_{j,i}\in M_j \cI$ and $\mu_i\in \phi_1M_i\cI$,  we have that $t_j\in \gvf_1^2 M_j^2 \cI^2$ if and only if $\gvf_1+k \phi_1 M_i/M_j\leq \gvf_1^2$. Writing $\gd=\max\{M_i/M_j \mid 1\leq i\leq k,; k+1\leq j\leq k+l\}$, we have that postponed reduction in step 6 in Algorithm 2 works if
\beql{LEprj} \gvf_1+k\phi_1 \gd \leq \gvf_1^2. \eeql

For example, suppose that on the bottom layer, both the left and right RNS have $k_1$ moduli, and that the $\gre$-value $\gre_1$ is 1/2, the optimal value. Then on the layer on top of the bottom layer, we have that $U_1=k_1$, $\gvf_1=k_1/\gre_1=2k_1$ and $\phi_1=k_1+1-\gre_1=k_1+1/2$; moreover, all moduli will have approximately the same (very large) size, so $\gd\approx 1$. Now the necessary condition  (\ref{LEprj}) reduces to $2k_1+k((k_1+1/2)\gd \leq 4k_1^2$, or
\beql{LEkcond} k \leq 4k_1/\gd.\eeql
Similarly, we may replace step 14 in Algorithm 2 by\\
$s_i=G_{i,0}'\eta_0+ G_{i,k+1}' \eta_{k+1}+ \cdots G_{i,k+l}' \eta_{k+l}$;\\
$\xi_i=\cR_{(M_i,m)}(s_i)$,\\
where $G'_{i,j}=\red{G_{i,j}m}{M_i}$ for all $j$. Again, this will work provided that $t_i$ can be computed and satisfies $t_i\in \gvf_1^2M_i^2 \cI^2$. In a similar way, writing $\gd'=\max\{M_j/M_i \mid 1\leq i\leq k,; k+1\leq j\leq k+l\}$ and $\gw=\max_{1\leq i\leq k}M_0/M_i$, we have that postponed reduction in step 14 in Algorithm 2 works if
\beql{LEpri} \gw+l \phi_1 \gd' \leq \gvf_1^2. \eeql
Normally, $\gw\ll1$, hence with the same assumptions on the bottom level, we now find that the necessary condition (\ref{LEpri}) will certainly be satisfied if $l\leq 4k_1/\gd'$. 
\subsection{Some improvements}
The algorithm in Section~\ref{LSS-alg4} can be slightly improved. Indeed a careful choice of the representing constants $H_s$ and of the signs $S_i$ may allow to skip steps 2 and 10 in Algorithm 2. 

First, if we choose
\[H_{j}=\red{J_{j}^{-1}}{M_{j}}=\red{M'/M_j}{M_j}\] 
for $j=k+1, \ldots, k+l$, then $E_{j}=m$, hence $\eta_{j}\equiv \xi_{kj}\bmod M_{j}$ for $j=k+1, \ldots, k+l$; as a consequence, we may be able to  skip step (\ref{LE-etaj}) of the algorithm, that is, step 10 in Algorithm 2. A slight complication is that the range of the $\eta_j$ was smaller than that of the $\xi_j$: we have $\eta_j\in \phi_1 M_j \cI$, but $\xi\in \gvf_1 M_j \cI$.  As a consequence, the bound (\ref{LEpri}) required for postponed reduction should be replaced by the bound
\beql{LEprjmod}  \gw/\gvf_1+l \gd' \leq \gvf_1. \eeql
For example, if on the level above the bottom level we have that $\gvf_1=k_1/\gre_1$, then we can satisfy this bound by taking $\gre_1$ small enough. However, note that as a consequence of choosing a smaller $\gre$, the upper bound on moduli $N$ on the next level will get smaller. 

Similarly,  if we choose 
\[ K_i=\red{-NS_i(M/M_i)}{M_i}\]
then $C_i=m$, and hence $\mu_i\equiv \chi_i \bmod M_i$ for $i=1, \ldots, k$. In that case, we may be able to  skip step 2 of Algorithm 2.
In the full Montgomery multiplication algorithm, we would have $K_i=H_i^2m$ after algorithm 1; so for the improvement, we would require that 
\[H_i^2\equiv  -N(M/M_i)S_im^{-1} \bmod M_i.\]
This choice is only available if 
the right-hand side is a square modulo $M_i$, but this could be achieved by choosing $S_i=1$ if it is a square and $S_i$ a quadratic non-residue if it is a non-square. In addition, we need the $S_i$ to be small in order to get a good upper bound on $u$. In the case of symmetric residues, one attractive choice is to take every $M_i$ prime with $M_i\equiv 3 \bmod 4$, so that $-1$ is a non-square modulo $M_i$ (such a restriction on the top-layer moduli is almost for free); Then we can choose $S_i=1$ or $S_i=-1$ to ensure that $-N(M/M_i)S_im^{-1}$ a square. 
Remark that the upper bound $U$ on $u$ will not be influenced by this choice of the $S_i$. On the other hand, in the case of symmetric residues, we should take $S_i$ positive but small; in that case we only take moduli $M_i$ for which there exists a small positive non-square when $-N(M/M_i)S_im^{-1}$  is a non-square. 

Again, 
even if $\mu_i\equiv \chi_i \bmod M_i$ for $i=1, \ldots, k$,  the $\mu_i$ have expansion $\phi_1$ while the $\chi_i$ have larger expansion $\gvf_1$. Similarly, the bound (\ref{LEprj}) required for postponed reduction should be replaced by
\beql{LEptimod}  1+l \gd \leq \gvf_1. \eeql


%
\section{\label{LS-comp}Complexity estimates and optimization} 
%
Consider a three-layer RNS system to implement exponentiation modulo $N$ by repeated Montgomery multiplications, consisting of a bottom, layer-1 RNS with moduli $m_0; m_1, \ldots, m_{k_1}; m_{k_1+1}, \ldots, m_{k_1+l_1}$, a middle, layer-2 RNS with moduli $M_0; M_1, \ldots, M_{k}; M_{k+1}, \ldots, M_{k+l}$, and a top, layer-3 RNS consisting of a single modulus $N$. We now give an estimate for the number of  simple operations required for a Montgomery multiplication modulo $N$, where a simple operation is a modular operation for the ``small'' moduli $m_i$ or for the redundant modulus $M_0$ (we assume that all such operations are approximately equally costly). 
Here we assume that on level 2 and 3, a Montgomery multiplication is implemented by Algorithm2 1 and~2, where steps 6 and 14 are done by postponed reduction.
Let $A_i, M_i, \cR_i, \cM_i$ denote the number of simple operations required for an addition, a multiplication, a Montgomery reduction, and a Montgomery operation on layer $i$, respectively.  
Then $A_1=M_1=\cM_1=1$
and $\cR_1=0$ since all arithmetic on the bottom layer is exact. Moreover, we have that $\cM_{t+1}=M_{t+1}+\cR_{t+1}$,  $A_{t+1}= (k_t+l_t)A_t+1$, and  $M_{t+1}=(k_{t}+l_t) \cM_{t} +1$ for $t\geq 1$. Finally, carefully counting the contributions of the various steps in Algorithm 2 gives that
\beqa R_{t}&=& k_t\cM_{t-1} +(2k_t+1)
	+l_t((k_t+1)M_{t-1}+k_tA_{t-1})\\
	&&+l_t\cR_{t-1} 
	+l_t\cM_{t-1}+(2l_t +1)
	+k_t((l_t+1)M_{t-1}\\
       &&+l_tA_{t-1}) 
	+k_t \cR_{t-1} \\
&=&  (k_t+l_t)\cM_{t-1}+2(k_t+l_t+1)\\
       && +(2k_tl_t+k_t+l_t)M_{t-1}\\
       && +2k_tl_tA_{t-1}+(k_t+l_t)\cR_{t-1}
\eeqa
for $t\geq2$.
It can be shown that in an optimal system, $k_t\approx l_t$ for $t=1,2$; assuming this, we find that $A_2\approx M_2=2k_1$, $\cR_2\approx 4k_1^2$, $\cM_2\approx 4k_1^2+2k_1$, 
and  on the top layer, we have  $M_3=2k\cM_1\approx 8kk_1^2$, $\cR_3\approx 16kk_1^2+8k^2k_1$. So the total number $\cM_3$ of simple operations  required for a Montgomery multiplication modulo $N$ satisfies
\beql{LEcM2}\cM_3\approx   24kk_1^2+8k^2k_1.\eeql
  
%
%

Now suppose that we employ table lookup to implement the simple operations, using tables of size $t\times t$ bits (so $m_i\leq 2^t$ for all $i$), and we want to be able to handle RSA moduli $N$ up to $b$ bits.
So for the upper bound $B_0$ for the layer-1 moduli, we have $B_0=2^t$. Then $m=m_1\cdots m_{k_1}\approx 2^{k_1t}$, so for the upper bound $B_1=m\gre_0(1-\gre_0)/U_0$ for the layer-2 moduli $M_s$, we have $B_1\approx 2^{tk_1}$. Similarly, $M=M_1\cdots M_{k}\approx 2^{tk_1k}$, so for the upper bound $B_2=M(\gre_1(1-\gre_1)/U_1$ on the modulus $N$, we have $B_2\approx 2^{tk_1k}$.  So we conclude that $b\approx tk_1k$, that is, $k\approx b/(k_1t)$. Using this value for $k$ in the expression for $\cM_2$ as given in 
(\ref{LEcM2}) and minimizing for $k_1$ results in minimizing value $k_1\approx \sqrt{b/(3t)}$, and minimum
\beql{LEtopsmin} \cM_2\approx 16\sqrt{3}\left(b/t\right)^{3/2}.\eeql

For example, if we want $t=8$ (byte-based tables) and $b=2048$ (modular arithmetic for 2048-bits RSA moduli), then for the optimal $k_1$, we find
\[ k_1\approx \sqrt{2048/24 } =\sqrt{84.333}\approx 9.\]
In practice, it turns out that the number of table operations is minimized when $k_1=9$ and $ k=32$.
%
\section{\label{LS-ex} An example}
To test and verify the algorithms in this paper, we have implemented in C++ a 2-layer RNS algorithm  for modular exponentiation 
with a $2048$-bits RSA modulus~$N$. Effectively, this is just a 3-layer RNS system with a top-layer consisting of a single $2048$-bits modulus. Note that  changing the RSA  modulus $N$ in the top-layer amounts to adapting some of the constants in Algorithm 2 for computing modulo $N$ on the top-layer with the RNS on the middle layer; the 2-layer RNS below remains unchanged.

For simplicity, we  used standard residues (so $\cI=[-1+e, e) =[0,1)$ with $e=1$), and we employed tables of size $8 \times 8 $ bits for the required modular arithmetic. So 
we use moduli of size at most $256$ on the bottom layer, and assumption $\cA(B_0=2^8, 1, \cI=[0,1), \gvf_0=1, \phi_0=1)$ holds.  For the bottom layer, it turns out that it is optimal to 
have one redundant modulus and 18 further moduli.
For these small moduli, we take the primes 
\[191,193,197,199,211,223,227,229,233,239,241,251,\] 
which are the 12 largest primes less than 256, and the composite numbers 
\[256=2^8, \ 253=11\cdot23, \  249=3\cdot 83, \  247=13\cdot 19,\]
\[ \ 235=5\cdot 47, \ 217=7\cdot31,\]
which are the largest numbers of the from $p^ia$ with $a>13$ prime, and which produces the largest attainable product for any list of 18 mutually co-prime numbers of size at most 256. Note that $255= 3\cdot5\cdot 17$ is a worse choice for both 3 and 5, similarly $245=5\cdot 7^2$ is a worse choice for both 5 and 7; the choices for 2, 11, and 13 are evidently optimal. Note that, as a consequence, the small moduli involve as prime factors all primes from 191 to 251 together with the primes $2,3,5,7,11,13,19,23,31,47,83$. So for the bottom-layer redundant modulus, we can take $m_0=17$.

Note that the bottom layer moduli have expansion coefficients $\gvf_0$ and $\phi_0$ for which $\gvf_0=\phi_0=1$.
Let $m$ and $m'$ denote the dynamical range of the left RNS $\cB_1$ and the right RNS $\cB_1'$, respectively. Let $\gre_1=1/2$ (the optimal choice), the best partition of these 18 bottom moduli such that $m'\geq (1-\gre_1)m$ with $m$ maximal turns out to result in 
a left RNS $\cB_1=(256, 251, 249, 247, 241, 239, 235, 199, 197)$, of size $k_1=9$ and with 
$m=2097065983013254306560$,
and 
a right RNS  
$ \cB'_1=(191, 193, 211, 217, 223, 227, 229, 233, 253)$, of size $l_1=9$ with  $m'= 1153388216560035715721$. Note that $m_0=17> k_1=9=l_1$ as required. Let $U_1=k_1 \phi_0=k_1=9$, and put $B_1=\lfloor  \gre_1 (1-\gre_1)m /U_1\rfloor=57669314532864493430$, $\gvf_1=U_1/\gre_0=18$, and $\phi_1=U_1+1-\gre_0=9.5$. Then the given RNS $\{m_0\}\cup \cB_1\cup \cB'_1$ can be used to realize assumption $\cM(B_1, m, \cI, \gvf_1, \phi_1)$. 

Now the choice of the large moduli $M_s$ on the middle layer is more or less automatic: if we need $k$ moduli for the left RNS $\cB$, we simply take the $k$ largest primes below $B_1$; then with $\gre=1/2$, in order to realize $M'\geq (1-\gre)M$ we can take $l=k$ and and take the next $l$ largest primes for the right RNS~$\cB'$. 
We want this layer to realize assumption $\cM(B,M,\cI,\gvf, \phi)$ with $B\geq2^{2048}$ so that we can handle RSA moduli $N$ on the top layer with up to 2048 bits.  
It turns out that in order to have $B=M\gre(1-\gre)/U$ large enough,
we need to take $k=32$
lower primes below $B_1$.
%
%
 For the redundant modulus, we can take $M_0=m_0m_j$ for some $j>k$ (in our program, we took $M_0=17\times 253$). It turns out that the parameters allow postponed reduction (see \Sec{LSpp}), which greatly increases the efficiency of the program.
\bre{LR-smods}\rm
It is possible to build a multi-layer RNS with unbounded dynamical range starting with a bottom layer of moduli of at most 4 bits. Indeed, take $\cB=(16, 15)$, $\cB'=(13,11)$, and $M_0=7$. Assuming exact arithmetic for all 4-bit moduli, we have $\cA(B_1=8, m=1, \cI=[-1/2, 1/2), \gvf_1=1, \phi_1=1)$, hence according to \Tm{LTbounds} with $k=l=2$, $M=16.15=240$, $M'=13.11=143$, and taking  $\gre=1/2$,  we have $U=1$ and $B=M.(1/4)/1=80$, and hence we can use this RNS to realize assumption $\cA(B=80, M=240, \cI=[-1/2, 1/2), \gvf=2, \phi=3/2)$.  From here on it is easy  to further increase the dynamical range by adding further (virtual)  layers.
\ere
\section{Conclusions}
We have 
presented an improved Bajard-Imbert-type full RNS algorithm that can also operate on inputs represented by pseudo-residues. Using this algorithm, we have developed 
a multi-layer Residue Number System (RNS) 
that is capable of implementing  modular addition, subtraction and  multiplication
 for very large moduli by only using actual arithmetic for a fixed set of moduli. If the moduli of this fixed set are sufficiently small,
 the method allows for a fully table-based implementation. 
In contrast to digit-based implementations of modular operations for large moduli, our
method allows for a massively parallel implementation and is completely carry-free, thus thwarting potential attacks exploiting such carries, e.g., with side-channel analysis or in a white-box cryptography context. 

Our system may  be considered as a method to provide a given, fixed RNS with a very large dynamical range.
To illustrate the method, we have described a 2-layer RNS system that can be used to implement an RSA exponentiation 
by adding the desired RSA modulus on top in a third layer. 
%
The system employs 19 moduli of 8-bits each in  the bottom layer and can be  used to implement 
an RSA exponentiation for 2048-bits  RSA moduli   with all the required arithmetic done by table look-up, using 19 modular addition tables and 19 modular multiplication tables,
 each of these 38 tables having size $2^8\times 2^8\times 8$ bits, with one modular multiplication taking approximately 160,000 table look-ups.  We further observed that in order to change the RSA modulus,  only some constants for computing on the top layer with moduli on the middle layer need to be updated. This update need not be computed in a secure manner and hence can be done quickly.

Our straightforward (non-parallelized) C++ program implementation of this RSA  exponentiation method with table-lookup  takes approximately 0.3 second on a HP Elitebook Folio 9470m  laptop to realize 500-bit modular exponentiation for a 2048-bits RSA modulus. So the security wish to remove all carries in the arithmetic can be achieved with a implementation operating at an acceptable speed.
%

%
%
%
\bibliographystyle{IEEEtran}
\bibliography{IEEEabrv,hdlh-mf}

\end{document}